\documentstyle{article}
\title{
\begin{flushright}
{\bf\normalsize   COLO-HEP-297 \\ UB-ECM-PF-93/3 }  \\
\end{flushright}
\vskip 10pt
\bf Steiner Variations on Random Surfaces
}

\author{ {\it C.F. Baillie} \\
         Physics Dept. \\
         University of Colorado\\
         Boulder, CO 80309\\
         USA\\
         \\
         {\it D. Espriu} \\
         DECM\\
         Universitat de Barcelona\\
         Diagonal 647\\
         08028 Barcelona\\
         Spain\\
         \\
         {\it D.A. Johnston} \\
         Dept. of Mathematics\\
         Heriot-Watt University\\
         Riccarton\\
         Edinburgh, EH14 4AS\\
         Scotland}

\begin{document}
  \maketitle
%-----------------------------------------------------------------------
                      {\Large
                      \begin{abstract}
%-----------------------------------------------------------------------
%
Ambartzumian et.al. suggested that the modified Steiner action functional had
desirable
properties for a random surface action. However, Durhuus and Jonsson pointed
out that such an action
led to an ill-defined
grand-canonical partition function and suggested that the addition of an area
term might improve
matters. In this paper we investigate this and other related actions
numerically for dynamically triangulated random surfaces and compare the
results with
the gaussian
plus extrinsic curvature actions that have been used previously.

%
%-----------------------------------------------------------------------
                        \end{abstract} }
%-----------------------------------------------------------------------
%
  \thispagestyle{empty}
%
%***********************************************************************
%
  \newpage
%
%-----------------------------------------------------------------------
                  \pagenumbering{arabic}
%-----------------------------------------------------------------------

There has been considerable recent interest in the theory and simulation of
random surfaces,
inspired both by string theory and the study of membranes in solid state
physics.
The Polyakov partition function \cite{1} for a string embedded in euclidean
space with
a fixed intrinsic area worldsheet discretizes to
\begin{equation}
Z = \sum_{T} \int \prod_{i=1}^{N-1} d X_i^{\mu} \exp (- S_g),
\label{e2}
\end{equation}
where $\sum_{T}$ is a sum over different triangulations of the worldsheet with
the same number of nodes $N$
which is the discrete analog of the path integration over the intrinsic metric
in the continuum.
The discretized action, $S_g$, is just a simple gaussian
\begin{equation}
S_g = {1 \over 2} \sum_{<ij>} (X_i^{\mu} - X_j^{\mu})^2,
\label{e3}
\end{equation}
where the variables $X_i$ live on the nodes of the triangulation and the sum
$<ij>$ runs over its edges.
To a solid state physicist this would be the action for a fluid
(because of the sum over triangulations) membrane with gaussian interactions
and no
self-avoidance.
The difficulties encountered in analytical calculations for string theories
\cite{2}
in physical dimensions are mirrored in simulations of equ.(\ref{e2})
which generate very crumpled surfaces from which it is impossible to take a
continuum limit \cite{3}.
Variations on the gaussian action, such as using the sum of the area of the
triangles making up the surface
or the sum of their edge lengths, also fail to generate a smooth continuum
limit \cite{4},\cite{5}.

The addition of an extrinsic
curvature term, or ``stiffness'', to the gaussian action which
originally arose in the context
of biological membranes and QCD \cite{6} has been mooted as a possible
resolution of these problems.
One discretization of this has shown particular promise,
\begin{equation}
S_e = \sum_{\Delta_i, \Delta_j} ( 1 - n_i \cdot n_j),
\label{e4}
\end{equation}
where the normals $n_i, n_j$ are on adjacent triangles $\Delta_i, \Delta_j$.
The strategy is to
examine the phase structure of the
gaussian plus extrinsic curvature action $S_g + \lambda S_e$
(henceforward GPEC) as the coupling $\lambda$ is varied
to see if any continuous transitions are present at which one might define
a non-trivial continuum theory. The case of fixed, or ``tethered'', surfaces
where there is no sum over triangulations offers some hope because there does
appear to
be a second order transition between a low $\lambda$ crumpled phase and a large
$\lambda$
smooth phase \footnote{These have been proposed as models of polymerized
membranes in solid state
physics \cite{6b}} for these when there is no self-avoidance \cite{6a}. Initial
work on
dynamical surfaces with modestly sized meshes tended to support a similar
picture \cite{7},
with a peak in the specific heat that grew with mesh size. More recent
simulations
\cite{8},\cite{9} with higher statistics and larger meshes have cast doubt on
this because the
increase in the peak height tails off
with increasing mesh size. This is not in itself worrying since it is possible
that the
string tension and mass gap may still scale in such a way as to give a non
trivial continuum
limit, even at a higher order transition \cite{8}. However, it was pointed out
in \cite{9}
that the data might also be construed as indicating a rapid crossover rather
than a true
transition or even the presence of a low mass bound state like that in the $2d$
$O(3)$ model.

This rather confused picture for GPEC actions naturally
prompts the
question of whether discrete models with more clearly cut phase transitions
exist. It was suggested in
\cite{10} that an action based on the modified Steiner functional \cite{11}
might be a suitable alternative candidate as it possessed the nice geometrical
property
of being subdivision invariant
as well as retaining a desirable stiffening effect on the surfaces.
The action was
\begin{equation}
S_{steiner} = {1 \over 2} \sum_{<ij>} | X_i^{\mu} - X_j^{\mu} | \theta
(\alpha_{ij}),
\label{e4a}
\end{equation}
where
$\theta(\alpha_{ij}) = | \pi - \alpha_{ij} |$ and $\alpha_{ij}$ is the angle
between the
embedded neighbouring triangles with common link $<ij>$.
This is essentially a coarse discretization of the absolute value of the
trace of the second fundamental form
of the surface.
However it was observed in
\cite{12} that an action of this form ran into problems with the entropy of
vertices in
smooth configurations and failed to give a well-defined grand canonical
partition function.
Our previous {\it microcanonical} simulation of such an action \cite{13} had
shown that it produced smooth surfaces
but in this case of course it is not clear how to obtain the continuum limit,
where
the problems of \cite{12} would presumably resurface.

We had observed in \cite{13} that varying the coupling $\lambda$ in an action
of the form
\begin{equation}
S_1 = {1 \over 2} \sum_{<ij>} | X_i^{\mu} - X_j^{\mu} | + \lambda \sum_{<ij>} |
X_i^{\mu} - X_j^{\mu} |
\theta (\alpha_{ij})
\label{e5}
\end{equation}
would allow one to employ the approach used in our earlier simulations of GPEC
actions,
namely hunting for continuous transitions at which to define the continuum
theory.
We chose the first, edge-length, term in $S_1$ because of its simplicity and
similarity to the
second Steiner term.
The authors of
\cite{12} pointed out that an subdivision invariant action of the form
\begin{equation}
S_2 = \sum_{\Delta} |\Delta | + \lambda \sum_{<ij>} | X_i^{\mu} - X_j^{\mu} |
\theta (\alpha_{ij}),
\label{e6}
\end{equation}
where $|\Delta |$ is the area of a triangle, might cure the entropy problems of
$S_{steiner}$ by
virtue of the first term being coercive enough to compete with the entropy of
smooth surfaces.
$S_1$ might also be expected to effect a cure for the same reason, but it is no
longer
subdivision invariant. It is not clear that this is a prerequisite for a
well-behaved action in any
case as the GPEC action does not share this property either. If we do not
insist on subdivision invariance
we could also envisage combining a gaussian term with $S_{steiner}$
\begin{equation}
S_3 = {1 \over 2} \sum_{<ij>} ( X_i - X_j )^2 + \lambda \sum_{<ij>} | X_i^{\mu}
- X_j^{\mu} | \theta (\alpha_{ij}).
\label{e7}
\end{equation}

The arguments presented in \cite{10} suggested that $S_{steiner}$ was
essentially a
discretization of the absolute value of the trace of the second fundamental
form,
at least for surfaces embedded in three dimensions. Making use of
the original definition of the trace of the second fundamental form
\begin{equation}
K = K^a_a  \simeq \partial^a X^{\mu} \cdot \partial_a n_{\mu}
\label{e13}
\end{equation}
we could also attempt a direct discretization in the three dimensional case.
Various possibilities are indicated in
Fig.1. One could take
\begin{equation}
 |K | = \sum_{<ij>} | ( X_i - X_j) \cdot ( n_1 - n_{2,3}) |.
\label{e14}
\end{equation}
where the subscript $2,3$ denotes a choice of either normal $n_2$ or $n_3$
in the discretization
\footnote{Choosing $X_j-X_k$ instead of $X_i - X_j$ gives identical results}
of $\partial n$.
If we pick $n_2$ it disappears in the sum because
it is perpendicular to $X_i - X_j$, whereas a coarser
discretization of the derivative given by choosing $n_3$ retains both of the
normals $n_1$ and $n_3$.
We can then pair these alternative
discretizations with either edge-length, area or gaussian terms to give the
further possibilities
\begin{eqnarray}
S_4 &=& {1 \over 2} \sum_{<ij>} | X_i^{\mu} - X_j^{\mu} | + \lambda \sum_{<ij>}
| K | \nonumber \\
S_5 &=& \sum_{\Delta} |\Delta | + \lambda \sum_{<ij>} | K | \nonumber \\
S_6 &=& {1 \over 2} \sum_{<ij>} ( X_i - X_j )^2 + \lambda \sum_{<ij>} | K | .
\label{e8}
\end{eqnarray}

In this paper we
carry out a qualitative exploration of the phase structure of $S_1, S_2, S_3$,
$S_4$, $S_5$ and $S_6$ to see if there
is any justification for expecting a well-behaved theory.
In order to do this we measured
the standard repertoire of intrinsic and extrinsic variables.
We included a local factor in
the measure for
compatibility with our earlier simulations which can be exponentiated to give
\begin{equation}
S_m = { d \over 2 } \sum_i \log ( q_i ),
\label{e10}
\end{equation}
where $q_i$ is the number of neighbours of point $i$, and $d=3$ dimensions.
If we denote the general form of actions $S_1,...,S_6$ by $S_{\alpha} + \lambda
S_{\beta}$
we thus simulated $S_{\alpha} + \lambda S_{\beta} + S_m$.
We measured $<S_m>$ and the mean maximum number of neighbours $<max(q_i)>$ to
get some idea
of the behaviour of the intrinsic geometry.
The extrinsic geometry
was observed by measuring $<S_{\beta}>$ and its
associated specific heat
\begin{equation}
C_{\beta} = {\lambda^2 \over N} \left( < S_{\beta}^2 > - < S_{\beta} >^2
\right)
\label{e11}
\end{equation}
as well as the gyration radius $X2$, a measure of the mean size of the
surface as seen in the embedding space,
\begin{equation}
X2 = { 1 \over 9 N (N -1)} \sum_{ij} \left( X^\mu_i - X^\mu_j \right)^2
q_i q_j.
\label{e12}
\end{equation}
Note that the sum for $X2$ is over all pairs of $X$'s on the mesh.
The expectation value of the first term in the action $<S_{\alpha}>$ was also
measured for completeness,
although unlike the GPEC actions we can no longer use its value
to confirm that equilibration has occurred.

The simulation used a Monte Carlo procedure
which we have described in some detail elsewhere \cite{15}. It first goes
through the mesh moving the $X$'s, carrying out a Metropolis
accept/reject at each step, and then goes through the mesh again
carrying out the ``flip'' moves on the links, again applying a
Metropolis accept/reject at each stage. The entire procedure
constitutes a sweep. Due to the correlated nature of the
data, a measurement was taken every tenth sweep and binning
techniques were used to analyse the errors. We carried out 10K
thermalization sweeps  followed by 30K or 50K measurement sweeps for
each data point. The acceptance for the $X$ move
was monitored and the size of the shift was adjusted to maintain an
acceptance of around 50 percent. The acceptance for the flip move was
also measured, but in this case there is nothing to adjust, so as for
GPEC actions this dropped with increasing $\lambda$ (but
was still appreciable even for quite large $\lambda$).

We now move on to consider the numerical results in more detail
for action $S_1$ (edge-length plus Steiner term)
on a 72-node surface. We can see from Table.1
\vskip 10pt
\hoffset=0truein
\voffset=0truein
\centerline{
\hbox{\vbox{ \tabskip=0pt \offinterlineskip
\def\tablerule{\noalign{\hrule}}
\halign{\strut#& \vrule#&
\hfill#\hfill &\vrule#& \hfill#\hfill &\vrule#&
\hfill#\hfill &\vrule#& \hfill#\hfill &\vrule#&
\hfill#\hfill &\vrule#& \hfill#\hfill &\vrule#&
\hfill#\hfill &\vrule#& \hfill#\hfill &\vrule#
\tabskip=0pt\cr
\tablerule
\omit&height2pt& \omit&height2pt&
\omit&height2pt& \omit&height2pt&
\omit&height2pt& \omit&height2pt&
\omit&height2pt& \omit&height2pt&
\omit&\cr
&&\enskip $\lambda$ \enskip
&&\enskip sweeps \enskip
&&\enskip $S_{\alpha}$ \enskip
&&\enskip $S_m$ \enskip
&&\enskip $S_{\beta}$ \enskip
&&\enskip $C_{\beta}$ \enskip
&&\enskip $X2$ \enskip
&&\enskip $max(q_i)$ \enskip
&\cr
\omit&height2pt& \omit&height2pt&
\omit&height2pt& \omit&height2pt&
\omit&height2pt& \omit&height2pt&
\omit&height2pt& \omit&height2pt&
\omit&\cr
\tablerule
\omit&height2pt& \omit&height2pt&
\omit&height2pt& \omit&height2pt&
\omit&height2pt& \omit&height2pt&
\omit&height2pt& \omit&height2pt&
\omit&\cr
&& 0.000 && 30K &&   213.08(0.20) &&   119.67(0.01) &&   382.66(0.35) &&
0.00(  0.00) &&     4.21(0.04) &&    16.31(0.02) &\cr
&& 0.500 && 30K &&   120.48(0.16) &&   120.37(0.01) &&   185.11(0.15) &&
0.64(  0.01) &&     1.64(0.02) &&    15.57(0.02) &\cr
&& 1.000 && 30K &&    88.53(0.12) &&   120.77(0.01) &&   124.43(0.08) &&
1.13(  0.01) &&     0.98(0.01) &&    15.21(0.02) &\cr
&& 1.500 && 30K &&    71.80(0.12) &&   121.10(0.01) &&    94.16(0.06) &&
1.40(  0.01) &&     0.71(0.01) &&    14.89(0.02) &\cr
&& 2.000 && 30K &&    61.43(0.16) &&   121.36(0.01) &&    75.84(0.06) &&
1.64(  0.02) &&     0.58(0.01) &&    14.60(0.03) &\cr
&& 2.500 && 30K &&    54.40(0.19) &&   121.61(0.02) &&    63.45(0.05) &&
1.81(  0.02) &&     0.50(0.01) &&    14.35(0.03) &\cr
&& 3.000 && 30K &&    49.91(0.50) &&   121.86(0.04) &&    54.34(0.12) &&
2.03(  0.10) &&     0.47(0.03) &&    14.09(0.06) &\cr
&& 3.250 && 30K &&    50.68(1.66) &&   122.12(0.10) &&    49.85(0.44) &&
2.31(  0.14) &&     0.60(0.10) &&    13.69(0.16) &\cr
&& 3.500 && 30K &&    83.55(1.04) &&   124.09(0.02) &&    37.23(0.13) &&
1.90(  0.11) &&     2.43(0.06) &&    10.45(0.04) &\cr
&& 4.000 && 30K &&    81.87(0.56) &&   124.21(0.01) &&    32.84(0.08) &&
1.58(  0.05) &&     2.34(0.03) &&    10.27(0.02) &\cr
&& 4.500 && 30K &&    79.34(0.61) &&   124.25(0.00) &&    29.72(0.08) &&
1.39(  0.03) &&     2.19(0.04) &&    10.20(0.01) &\cr
&& 5.000 && 30K &&    76.99(1.09) &&   124.25(0.00) &&    27.50(0.15) &&
1.35(  0.03) &&     2.08(0.07) &&    10.20(0.01) &\cr
\omit&height2pt& \omit&height2pt&
\omit&height2pt& \omit&height2pt&
\omit&height2pt& \omit&height2pt&
\omit&height2pt& \omit&height2pt&
\omit&\cr
\tablerule
}}}}
\smallskip
\centerline{Table 1}
\centerline{Results for $S_1$, N=72}
\vskip 5pt
\noindent
that there does, indeed,
appear to be a transition to a smooth phase at large $\lambda$, as inspection
of
Figs.2,3 confirms. We can see in Fig.2 that there is a peak in the associated
specific heat
$C_{\beta}$ at around $\lambda=3.25$ where we also observe a sharp jump in
the measured size of the surface as given by $X2$ which is plotted in Fig.3.
The smooth nature of the large $\lambda$ phase is corroborated by the snapshot
of a surface at $\lambda=5$ in Fig.4.
For small $\lambda$ the surfaces are crumpled as can be seen in Fig.5.
The behaviour in the crumpled phase is, however, different
from that seen with GPEC actions. Firstly, at $\lambda=0$ the surfaces
are larger than those seen with a purely gaussian action,
presumably because the edge length action is less confining than the gaussian
action.
Secondly, and rather strikingly, $X2$ {\it decreases} with
increasing $\lambda$ up to the putative phase transition point, where it
increases again.
This can be understood by noting that the Steiner term, $S_{\beta}$, is not
scale invariant like the extrinsic curvature. There are thus
two ways of decreasing the value of $S_{\beta}$, one can either make
the surface smaller to decrease the $|X_i - X_j|$ terms, or make the surface
smoother to decrease the $\theta$ terms. It is evident that the first
occurs for $\lambda$ up to the phase transition, as can be confirmed by
looking at $S_{\alpha}$, which is a measure of the sum of $|X_i - X_j|$ terms -
this decreases. The behaviour of $X2$ in the crumpled phase
appears to be a generic feature of all of the actions containing Steiner
terms as it is also observed for $S_2$ (area $+$ Steiner) and $S_3$ (gaussian
$+$ Steiner) actions.
The Steiner term $S_{\beta}$ appears to correlate well with
the extrinsic curvature, as can be seen in Table.1, where its value is high in
the
crumpled low $\lambda$ phase (where the extrinsic curvature, which we have not
tabulated, is also high) and drops off as $\lambda$ increases and the surfaces
become smoother.
The behaviour of the intrinsic observables $S_m$ and $max(q_i)$ is very similar
to that of the GPEC
actions: $S_m$ increases smoothly through the phase transition and $\max(q_i)$
decreases,
indicating that the {\it intrinsic} geometry becomes slightly more regular at
larger $\lambda$.

$S_2$ (area plus Steiner term) gives the results in Table.2.
Again we have a smooth phase at large $\lambda$ with a transition to
a crumpled phase at lower $\lambda$. Snapshots of surfaces in the smooth phase
look similar
to that produced by $S_1$ in Fig.4 and the low $\lambda$ phase is, if anything,
even spikier than that of $S_1$. In fact, at $\lambda=0$ the surfaces appear
to be unstable to the formation of very large spikes, appearing as a collection
of thin whiskers emanating from a central point. We have not included
the values for $\lambda=0$ in Table.2 and Figs.2,3 as $X2$ and $S_{\beta}$ are
very large.
As we can see from Table.2
\vskip 10pt
\hoffset=0truein
\voffset=0truein
\centerline{
\hbox{\vbox{ \tabskip=0pt \offinterlineskip
\def\tablerule{\noalign{\hrule}}
\halign{\strut#& \vrule#&
\hfill#\hfill &\vrule#& \hfill#\hfill &\vrule#&
\hfill#\hfill &\vrule#& \hfill#\hfill &\vrule#&
\hfill#\hfill &\vrule#& \hfill#\hfill &\vrule#&
\hfill#\hfill &\vrule#& \hfill#\hfill &\vrule#
\tabskip=0pt\cr
\tablerule
\omit&height2pt& \omit&height2pt&
\omit&height2pt& \omit&height2pt&
\omit&height2pt& \omit&height2pt&
\omit&height2pt& \omit&height2pt&
\omit&\cr
&&\enskip $\lambda$ \enskip
&&\enskip sweeps \enskip
&&\enskip $S_{\alpha}$ \enskip
&&\enskip $S_m$ \enskip
&&\enskip $S_{\beta}$ \enskip
&&\enskip $C_{\beta}$ \enskip
&&\enskip $X2$ \enskip
&&\enskip $max(q_i)$ \enskip
&\cr
\omit&height2pt& \omit&height2pt&
\omit&height2pt& \omit&height2pt&
\omit&height2pt& \omit&height2pt&
\omit&height2pt& \omit&height2pt&
\omit&\cr
\tablerule
\omit&height2pt& \omit&height2pt&
\omit&height2pt& \omit&height2pt&
\omit&height2pt& \omit&height2pt&
\omit&height2pt& \omit&height2pt&
\omit&\cr
&& 1.000 && 30K &&    35.23(0.09) &&   121.08(0.01) &&   142.72(0.02) &&
1.28(  0.00) &&     1.64(0.02) &&    14.90(0.02) &\cr
&& 1.250 && 30K &&    29.76(0.01) &&   121.34(0.00) &&   122.66(0.02) &&
1.50(  0.00) &&     1.47(0.00) &&    14.60(0.00) &\cr
&& 1.500 && 30K &&    25.79(0.04) &&   121.58(0.00) &&   107.23(0.01) &&
1.76(  0.01) &&     1.34(0.01) &&    14.40(0.01) &\cr
&& 1.750 && 30K &&    23.55(0.05) &&   121.88(0.01) &&    94.74(0.05) &&
1.95(  0.00) &&     1.43(0.02) &&    14.02(0.02) &\cr
&& 2.000 && 30K &&    24.58(0.71) &&   122.33(0.05) &&    82.04(0.64) &&
4.01(  0.43) &&     1.87(0.13) &&    13.39(0.10) &\cr
&& 2.250 && 30K &&    36.87(0.82) &&   123.45(0.04) &&    62.48(0.50) &&
6.25(  0.24) &&     3.80(0.12) &&    11.51(0.07) &\cr
&& 2.500 && 30K &&    41.70(0.23) &&   123.87(0.01) &&    51.70(0.12) &&
2.49(  0.11) &&     4.50(0.02) &&    10.73(0.02) &\cr
&& 3.000 && 30K &&    40.95(0.17) &&   123.99(0.00) &&    43.95(0.02) &&
1.37(  0.01) &&     4.50(0.02) &&    10.56(0.00) &\cr
&& 4.000 && 30K &&    34.63(0.11) &&   124.02(0.00) &&    35.89(0.03) &&
1.40(  0.02) &&     3.80(0.01) &&    10.52(0.00) &\cr
&& 5.000 && 30K &&    29.09(0.09) &&   124.00(0.00) &&    30.98(0.03) &&
1.60(  0.01) &&     3.22(0.01) &&    10.54(0.00) &\cr
\omit&height2pt& \omit&height2pt&
\omit&height2pt& \omit&height2pt&
\omit&height2pt& \omit&height2pt&
\omit&height2pt& \omit&height2pt&
\omit&\cr
\tablerule
}}}}
\smallskip
\centerline{Table 2}
\centerline{Results for $S_2$, $N=72$}
\vskip 5pt
\noindent
the $X2$ values for $S_2$ even with non-zero $\lambda$ are larger for a given
coupling than for $S_1$,
suggesting the spikiness persists away from $\lambda=0$.
We can also see from the table that $X2$ decreases at small $\lambda$
as does $S_{\alpha}$ for precisely the reasons indicated above
for $S_1$.
It is clear that
the transition is taking place at a lower $\lambda$ value than for $S_1$, as
both
the peak in the specific heat and the jump in $X2$ are at $\lambda \simeq
2.25$, rather than
$\lambda \simeq 3.25$. At first sight this might appear rather surprising
as we have just seen that the crumpled phase is spikier for $S_2$ rather than
$S_1$, so we might expect to have to work harder to escape from it and thus
have a phase transition at {\it larger} $\lambda$. However, the dimensions
of the area and edge length terms are different so we cannot directly attach
significance to the numerical value of the Steiner coupling.
We can see in Fig.1 that the peak in the specific heat for
$S_2$ appears to be considerably stronger than for $S_1$
\footnote{It is worth remarking that the sharpness of the peak in the specific
heat
should not be regarded as the criterion for a good candidate continuum theory
as models which do have second order transitions on fixed
lattices, such as Ising or Potts models, have higher order transitions on
dynamical lattices
with well behaved continuum limits. The GPEC action \cite{6a}and $S_1$ may be
similar.}.
%The error bars at the peak in Fig.2 are almost certainly underestimated for
%%both
%$S_2$ and $S_3$ below because of our modest statistics and bin sizes used in
%%estimating them.

$S_3$ (gaussian plus Steiner term) gives the results shown in Table.3.
where we can see that the behaviour is similar to $S_2$. Although $S_3$ is not
subdivision invariant
there is still a larger peak in the specific heat than for $S_1$ as can be seen
clearly in
Fig.2. The peak is at a larger $\lambda$ ($\simeq 3$) value than for $S_2$,
which is perhaps surprising as one might have expected
that it would be easier to escape from the crumpled phase of $S_3$ because it
is less ``spiky'' than that of $S_2$.
On the other hand, as can be seen in Fig.3, the $X2$ values of $S_3$ are more
similar to $S_1$ than $S_2$.
The characteristic decrease in $X2$ for
small $\lambda$ is also present for $S_3$.
Again the values for $\lambda=0$ are omitted, this time because they are
identical to the GPEC action at $\lambda=0$.

\vskip 10pt
\hoffset=0truein
\voffset=0truein
\centerline{
\hbox{\vbox{ \tabskip=0pt \offinterlineskip
\def\tablerule{\noalign{\hrule}}
\halign{\strut#& \vrule#&
\hfill#\hfill &\vrule#& \hfill#\hfill &\vrule#&
\hfill#\hfill &\vrule#& \hfill#\hfill &\vrule#&
\hfill#\hfill &\vrule#& \hfill#\hfill &\vrule#&
\hfill#\hfill &\vrule#& \hfill#\hfill &\vrule#
\tabskip=0pt\cr
\tablerule
\omit&height2pt& \omit&height2pt&
\omit&height2pt& \omit&height2pt&
\omit&height2pt& \omit&height2pt&
\omit&height2pt& \omit&height2pt&
\omit&\cr
&&\enskip $\lambda$ \enskip
&&\enskip sweeps \enskip
&&\enskip $S_{\alpha}$ \enskip
&&\enskip $S_m$ \enskip
&&\enskip $S_{\beta}$ \enskip
&&\enskip $C_{\beta}$ \enskip
&&\enskip $X2$ \enskip
&&\enskip $max(q_i)$ \enskip
&\cr
\omit&height2pt& \omit&height2pt&
\omit&height2pt& \omit&height2pt&
\omit&height2pt& \omit&height2pt&
\omit&height2pt& \omit&height2pt&
\omit&\cr
\tablerule
\omit&height2pt& \omit&height2pt&
\omit&height2pt& \omit&height2pt&
\omit&height2pt& \omit&height2pt&
\omit&height2pt& \omit&height2pt&
\omit&\cr
&& 0.500 && 30K &&    65.08(0.02) &&   120.33(0.00) &&   166.39(0.02) &&
0.38(  0.00) &&     1.24(0.00) &&    15.61(0.00) &\cr
&& 1.000 && 30K &&    44.86(0.03) &&   120.81(0.00) &&   123.70(0.01) &&
0.91(  0.00) &&     0.98(0.00) &&    15.15(0.00) &\cr
&& 1.500 && 30K &&    33.52(0.04) &&   121.23(0.00) &&    97.39(0.01) &&
1.31(  0.01) &&     0.82(0.00) &&    14.72(0.00) &\cr
&& 2.000 && 30K &&    26.84(0.06) &&   121.59(0.01) &&    79.62(0.01) &&
1.63(  0.00) &&     0.74(0.01) &&    14.36(0.01) &\cr
&& 2.500 && 30K &&    22.91(0.05) &&   121.97(0.01) &&    66.65(0.03) &&
2.04(  0.01) &&     0.71(0.01) &&    13.96(0.01) &\cr
&& 2.950 && 30K &&    30.14(0.45) &&   123.11(0.03) &&    51.72(0.28) &&
6.45(  0.20) &&     1.49(0.03) &&    12.22(0.06) &\cr
&& 3.000 && 30K &&    32.54(0.24) &&   123.32(0.02) &&    49.42(0.17) &&
6.77(  0.19) &&     1.69(0.02) &&    11.87(0.05) &\cr
&& 3.050 && 30K &&    36.73(1.16) &&   123.68(0.07) &&    45.95(0.65) &&
5.39(  0.43) &&     2.04(0.09) &&    11.24(0.14) &\cr
&& 4.000 && 30K &&    41.32(0.14) &&   124.31(0.00) &&    32.61(0.01) &&
1.41(  0.01) &&     2.48(0.01) &&    10.11(0.01) &\cr
&& 4.500 && 30K &&    38.61(0.09) &&   124.32(0.00) &&    29.81(0.01) &&
1.24(  0.01) &&     2.31(0.01) &&    10.09(0.00) &\cr
&& 5.000 && 30K &&    36.94(0.17) &&   124.31(0.00) &&    27.88(0.02) &&
1.30(  0.01) &&     2.21(0.01) &&    10.11(0.00) &\cr
\omit&height2pt& \omit&height2pt&
\omit&height2pt& \omit&height2pt&
\omit&height2pt& \omit&height2pt&
\omit&height2pt& \omit&height2pt&
\omit&\cr
\tablerule
}}}}
\smallskip
\centerline{Table 3}
\centerline{Results $S_3, N=72$}
\vskip 5pt
\noindent

We might naively expect $S_4,S_5,S_6$ to behave in a similar fashion to
$S_1,S_2,S_3$ respectively, but a glance
at Fig.6 and Fig.7 where we show snapshots of 72-node surfaces generated by
$S_4$ and $S_5$ respectively at $\lambda=5$ reveal that this is not the case.
It is clear for both of these that $S_{\beta}$ is failing to smooth out the
surfaces
at large $\lambda$, for $S_4$ (and $S_6$ too,
which we have not shown) because of large numbers of ``back to back'' triangles
and
for $S_5$ because it fails to prevent the surface degenerating into thin
whiskers,
similar to those observed for $S_2$ at $\lambda=0$.
In Table.4
\vskip 10pt
\hoffset=0truein
\voffset=0truein
\centerline{
\hbox{\vbox{ \tabskip=0pt \offinterlineskip
\def\tablerule{\noalign{\hrule}}
\halign{\strut#& \vrule#&
\hfill#\hfill &\vrule#& \hfill#\hfill &\vrule#&
\hfill#\hfill &\vrule#& \hfill#\hfill &\vrule#&
\hfill#\hfill &\vrule#& \hfill#\hfill &\vrule#&
\hfill#\hfill &\vrule#& \hfill#\hfill &\vrule#
\tabskip=0pt\cr
\tablerule
\omit&height2pt& \omit&height2pt&
\omit&height2pt& \omit&height2pt&
\omit&height2pt& \omit&height2pt&
\omit&height2pt& \omit&height2pt&
\omit&\cr
&&\enskip $\lambda$ \enskip
&&\enskip sweeps \enskip
&&\enskip $S_{\alpha}$ \enskip
&&\enskip $S_m$ \enskip
&&\enskip $S_{\beta}$ \enskip
&&\enskip $C_{\beta}$ \enskip
&&\enskip $X2$ \enskip
&&\enskip $max(q_i)$ \enskip
&\cr
\omit&height2pt& \omit&height2pt&
\omit&height2pt& \omit&height2pt&
\omit&height2pt& \omit&height2pt&
\omit&height2pt& \omit&height2pt&
\omit&\cr
\tablerule
\omit&height2pt& \omit&height2pt&
\omit&height2pt& \omit&height2pt&
\omit&height2pt& \omit&height2pt&
\omit&height2pt& \omit&height2pt&
\omit&\cr
&& 1.000 && 50K &&    70.48(0.01) &&    58.83(0.00) &&    34.62(0.01) &&
0.61(  0.00) &&     1.57(0.00) &&     4.13(0.00) &\cr
&& 2.000 && 50K &&    64.95(0.07) &&    58.69(0.00) &&    20.05(0.01) &&
1.10(  0.00) &&     1.39(0.01) &&     4.54(0.00) &\cr
&& 3.000 && 50K &&    66.59(0.06) &&    58.66(0.00) &&    12.78(0.01) &&
1.30(  0.00) &&     1.42(0.01) &&     4.66(0.00) &\cr
&& 4.000 && 50K &&    69.11(0.14) &&    58.70(0.00) &&     9.01(0.01) &&
1.19(  0.00) &&     1.55(0.01) &&     4.57(0.00) &\cr
&& 5.000 && 50K &&    69.86(0.16) &&    58.72(0.00) &&     6.98(0.00) &&
1.10(  0.00) &&     1.59(0.01) &&     4.52(0.00) &\cr
&& 7.000 && 50K &&    70.68(0.22) &&    58.72(0.00) &&     4.85(0.01) &&
1.02(  0.01) &&     1.61(0.01) &&     4.52(0.01) &\cr
\omit&height2pt& \omit&height2pt&
\omit&height2pt& \omit&height2pt&
\omit&height2pt& \omit&height2pt&
\omit&height2pt& \omit&height2pt&
\omit&\cr
\tablerule
}}}}
\smallskip
\centerline{Table 4}
\centerline{Results for $S_4$, N=36}
\vskip 5pt
\noindent
the numerical results for $S_4$ show that despite the obvious
roughness of the surfaces at large $\lambda$, $S_{\beta}$ which is supposed to
discretize
the absolute value of the trace of the second fundamental form, is decreasing
with increasing $\lambda$.
The numerical results for $S_5, S_6$ are similar to this - although $S_{\beta}$
decreases with increasing $\lambda$ the surfaces obviously remain rough.
We have
not pursued simulations of $S_4,S_5,S_6$ on larger surfaces because of the
obvious pathologies
seen in these results.

For the choice of $n_2$ in the discretization of $|K|$ in equ.(\ref{e14})
the nature of the pathology can be pinned down by looking at  $S_{\beta}$
{\it without} the modulus sign for the various surfaces. Summing
up the various contributions on the smooth surfaces generated by $S_1, S_2,
S_3$ gives
a much larger total than summing up the contributions on surfaces generated by
$S_4,S_5,S_6$
where we see a large degree of cancellation between terms with alternating
signs.
The reason for this is clear if we examine equ.(\ref{e14}): there is an
ambiguity in the direction of $n_1$ if it is forced to be perpendicular to $X_i
- X_j$.
It can be at either $+ 90^o$ or $- 90^o$ and the Monte Carlo will pick both
with equal probability, leading
to normals reversing direction from triangle to triangle and the non-smooth
surfaces that are observed
with $S_4,S_5,S_6$. This problem does not arise for $S_1,S_2,S_3$ where the
Steiner term is
minimized when the angle between adjacent triangles is $180^o$ and there is no
possibility of confusion
in the orientation of the normals.  Choosing $n_3$ in the discretization of
$|K|$ in equ.(\ref{e14}), which retains
both normals,
might be expected to give better results as it is more difficult to generate
crumpled configurations
with both $n_1$ and $n_3$  perpendicular to $X_i - X_j$. Nonetheless,
this alternate discretization also fails to produce a smooth phase.
Examining $S_{\beta}$ without the modulus sign again reveals a large degree of
cancellation due to
alternating signs in the surfaces generated by $S_4,S_5,S_6$ (which is not seen
for the smooth surfaces
produced by $S_1, S_2, S_3$).

The simulations described in this paper can be regarded as a qualitative study
of the feasibility of using alternatives to the GPEC action for random
surfaces. We have found
that $S_1$ and $S_3$ which are not subdivision invariant and $S_2$, which is,
all seem to
be good candidates for further exploration.
$S_4$, $S_5$ and $S_6$ on the other hand,
fail to discretize the trace of the second fundamental form properly.
It thus appears that the choice of ``stiffness'' term essentially determines
at least the qualitative behaviour
of the surface at the crumpling transition. Similar conclusions have been drawn
by one of the authors
for various versions of the extrinsic curvature combined with area, gaussian or
tethering
terms on fixed random surfaces \cite{16}. The crumpled phase for the various
actions, particularly for $S_2$ at very small $\lambda$, does display
variations.
The interesting question of whether the subdivision invariance of $S_2$
has an effect on the random surface model as it approaches the continuum limit
has not been investigated in this paper
because
we have made no
attempt to carry out finite size scaling for $S_1,S_2,S_3$ at the transition.
This would also allow a more quantitative comparison with the GPEC actions
and establish whether the properties of the transition for $S_1,S_2,S_3$  are
more clear cut than
those seen in \cite{8}, \cite{9}.
These issues, and the equally important question
of whether the string tension and mass gap scale, are currently being addressed
in a much larger simulation.

The comparison with the results of the GPEC action will also shed some light on
the still
murky problem of universality for random surfaces \cite{17}.
As we have seen, there is a wide choice of possible
terms and discretizations for a random surface action and the question of
whether
they lead to the same continuum theory has not been clearly answered. In the
longer run
the best approach may be to use the Monte Carlo renormalization group, perhaps
starting on rigid meshes
because of the computational complexity of dynamical meshes. This would then
allow
consideration of {\it all} possible terms up to a given weight. We are
currently exploring
the feasibility of this.

This work was supported in part by NATO collaborative research grant CRG910091
(CFB and DAJ)
and an Spanish/British Acciones Integradas grant (DE and DAJ).
CFB is supported by DOE under contract DE-FG02-91ER40672 and by AFOSR Grant
AFOSR-89-0422.
The computations were performed on workstations at the University of Colorado
and Heriot-Watt University, and on the Meiko Computing Surface at Heriot-Watt.
We would like to thank
R.D. Williams for help in developing initial versions of the dynamical
mesh code.

\vfill
\eject

\vfill
\eject

\centerline{\bf Figure Captions}
\begin{description}
\item[Fig. 1.]
The edges and normals that are involved in the discretization of the
the $S_{\beta}$ term in $S_4$, $S_5$ and $S_6$.
\item[Fig. 2.]
The specific heat $C_{\beta}$ for actions $S_1,S_2,S_3$.
\item[Fig. 3.]
The gyration radii $X2$ for actions $S_1,S_2,S_3$.
\item[Fig. 4.]
A snapshot of a mesh generated by $S_1$ with $\lambda=5$.
\item[Fig. 5.]
A snapshot of a mesh generated by $S_1$ with $\lambda=0$.
\item[Fig. 6.]
A snapshot of a mesh generated by $S_4$ with $\lambda=5$.
\item[Fig. 7.]
A snapshot of a mesh generated by $S_5$ with $\lambda=5$.
\end{description}
\vfill\eject

\end{document}